\documentclass[aps,prb,twocolumn,showpacs,amsmath,amssymb]{revtex4}
\usepackage{graphicx}
\usepackage{epstopdf}

\begin{document}

\title{Correlation effects in Ni 3$d$ states of LaNiPO}

\author {A.~V.~Lukoyanov,$^{1,2}$ S.~L.~Skornyakov,$^{1,2}$ J.~A.~McLeod,$^3$ M.~Abu-Samak,$^4$ R.~G.~Wilks,$^3$ 
E.~Z.~Kurmaev,$^1$ A.~Moewes,$^3$ N.~A.~Skorikov,$^1$ Yu.~A.~Izyumov,$^1$ L.~D.~Finkelstein,$^1$ 
V.~I.~Anisimov,$^1$ and D.~Johrendt$^5$}

\affiliation{$^1$Institute of Metal Physics, Russian Academy of Sciences--Ural Division,
620990 Yekaterinburg, Russia\\
$^2$Ural State Technical University--UPI, 620002 Yekaterinburg, Russia\\
$^3$Department of Physics and Engineering Physics, University of Saskatchewan, 
116 Science Place, S7N 5E2 Saskatoon, Canada\\
$^4$Physics Department, Al-Hussein Bin Talal University, P.O. Box 20, Ma'an, Jordan\\
$^5$Department Chemie und Biochemie der Ludwig-Maximilians-Universit\"at M\"unchen, 
Butenandtstrasse 5-13 (Haus D), 81377 M\"unchen, Germany}

\date{\today}

\begin {abstract}
The electronic structure of the new superconducting material LaNiPO 
experimentally probed by soft X-ray spectroscopy and theoretically calculated 
by the combination of local density approximation with Dynamical Mean-Field Theory 
(LDA+DMFT) are compared herein. We have measured the Ni $L_{2,3}$ X-ray emission (XES) 
and absorption (XAS) spectra which probe the occupied and unoccupied the Ni 3$d$ states, respectively. 
In LaNiPO, the Ni 3$d$ states are strongly renormalized by dynamical correlations 
and shifted about 1.5 eV lower in the valence band than the corresponding Fe 3$d$ states in LaFeAsO. 
We further obtain a lower Hubbard band at --9 eV below 
the Fermi level in LaNiPO which bears striking resemblance to the lower Hubbard band in the correlated oxide NiO, while no such band is observed in LaFeAsO. These results 
are also supported by the intensity ratio between the transition metal $L_2$ and 
$L_3$ bands measured experimentally to be higher in LaNiPO than in LaFeAsO, 
indicating the presence of the stronger electron correlations in the Ni 3$d$ states 
in LaNiPO in comparison with the Fe 3$d$ states in LaFeAsO. These findings 
are in accordance with resonantly excited transition metal $L_3$ X-ray emission spectra 
which probe occupied metal 3$d$-states and show the appearance of the lower Hubbard 
band in LaNiPO and NiO and its absence in LaFeAsO.
\end {abstract}


\maketitle

\section{Introduction} 

The superconductivity of the quaternary transition metal oxyphosphides LaFePO ($T_c$~=~3.2~K) 
and LaNiPO ($T_c$ = 3.0 -- 4.3 K) was discovered recently\cite{Kamihara06,Watanabe07,Tegel08}
although these compounds have 
been under study for more than ten years.~\cite{Zimmer95} In spite of the relatively low superconducting 
transition temperatures $T_c$, these materials are important because they triggered the extensive 
search for superconductivity in oxyarsenides $LnMe$AsO (where $Ln$ = La, Ce, Pr, Nd, Sm, Gd; 
$Me$ = 3$d$ metals). This search has proven to be remarkably successful, reaching $T_c$s of up to 
55 K\cite{Kamihara08,Wen08,Chen08,Ren08a,Ren08b,Liu08} and strong upper critical fields H$_{c2}$ 
of up to 100~T.~\cite{Senatore08}

Despite the focus on oxyarsenides, Ni and Fe oxyphosphides still attract attention\cite{Lebeque07,Zhang08,Liang07,Che08,Si08} 
because while they share the same bi-layer structure as $LnMe$AsO materials, the mechanism 
for superconductivity appears to be different. In particular the value of the total 
electron-phonon coupling constant $\lambda$ in LaFeAsO is much lower than in conventional 
electron-phonon coupling superconductors, for example, compare the $\lambda$ = 0.21 of 
LaFeAsO\cite{Boeri08} with the $\lambda$ = 0.44 of Al (where $T_c$ = 1.3 K for Al), and even the inclusion 
of multiband effects fails to explain the observed $T_c$ of 26 K.~\cite{Boeri09} For LaNiPO the coupling 
constant is more then two times higher ($\lambda$ = 0.58) and the superconducting properties 
can be described within the Migdal-Eliashberg theory.~\cite{Boeri09} Herein we interpret soft X-ray 
emission and absorption spectra of LaNiPO and LaFeAsO\cite{Kurmaev08b} and compare 
the measurements with our local density approximation with Dynamical Mean Field Theory (LDA+DMFT) electronic structure calculations to investigate 
the similarities and differences between these two types of superconductors. 
To assist in investigating these materials, electronic structure calculations  
of LaFePO\cite{Skornyakov10} ($T_c$ = 3.2 K\cite{Kamihara06}) 
and NiO\cite{Kunes07} were used. 

LaFeAsO, unlike LaNiPO and LaFePO, is not superconducting unless doping\cite{Li08,Dong08}
or high pressure\cite{Yang09,Chu09} is applied to suppress the magnetic transition 
temperature.~\cite{Chu09} Since these materials share the same basic ambient crystal 
structure and atomic 
constituents, yet exhibit different low-temperature properties, a basic study of the ambient 
electronic structure of these three materials is of interest, especially since the bulk 
electronic structure of these materials is insensitive to temperature or magnetic phase 
changes.~\cite{Yildirim09}

\section{Experimental and Calculation Details} 

Single crystals of LaNiPO were synthesized by heating a mixture of 375.0 mg
La (99.9\%, Smart Elements), 201.7 mg NiO (99.99\%, Sigma-Aldrich) and 83.6 mg P 
(red, 99\%, Sigma-Aldrich) with 2000 mg Sn (99.99\%, Alfa Aesar) in an alumina crucible, 
which was sealed in a silica tube under an atmosphere of purified argon. The sample 
was heated to 1173 K at a rate of 40 K/h, kept at this temperature for 10 days 
and slowly cooled down to room temperature at a rate of 3 K/h. The crucible 
was smashed and the tin bar dissolved in 6 M HCl at room temperature. The remaining sample 
consisted of single crystals of LaNiPO beside small amounts of LaNi$_2$P$_2$ (7\%), Ni$_2$SnP (4\%) 
and Ni$_3$Sn$_4$ ($<$ 1\%). Further attempts to optimize the synthesis conditions with regard 
to reaction temperature or duration were unsuccessful. Samples prepared directly from 
the starting material without tin flux yielded only small amounts of LaNiPO with LaNi$_2$P$_2$ 
as the main product. For details of preparation see~Ref.~\onlinecite{Tegel08}. 
The soft X-ray absorption and emission measurements of the metal $L_{2,3}$ edges 
were performed at the soft X-ray fluorescence endstation 
of Beamline 8.0.1 at the Advanced Light Source in the Lawrence Berkeley National Laboratory.~\cite{Jia95} 
The endstation uses a Rowland circle geometry X-ray spectrometer with spherical gratings 
and an area-sensitive multichannel detector. We measured the resonant and non-resonant Ni $L_{2,3}$ 
(3$d$,4$s$ $\to$ 2$p$ transition) X-ray emission spectra (XES) for LaNiPO. Additional non-resonant XES 
measurements of the Ni $L_{2,3}$ edges of Ni metal foil and NiO were obtained as reference standard. 
The instrumental resolving power (E/$\Delta$E) for emission measurements 
was about 10$^3$. The X-ray absorption spectra (XAS) were measured in total electron yield (TEY) 
mode for the Ni $L_{2,3}$ edges. The instrumental resolving power (E/$\Delta$E) for absorption 
measurements was about 5 $\times$ 10$^3$. All absorption spectra were normalized to the incident photon 
current using a highly transparent gold mesh in front of the sample to correct for intensity 
fluctuations in the incoming photon beam. The excitation energies for the Ni $L_{2,3}$ resonant 
X-ray emission spectra were determined from the XAS spectra and the energies were selected 
at the $L_3$ and $L_2$ thresholds.

Electronic structure calculations were performed within 
the pseudopotential plane-wave method PWSCF, as implemented in 
the Quantum ESPRESSO package.~\cite{PW} We used the generalized gradient 
approximation in the Perdew-Burke-Ernzerhof version\cite{Perdew96}
for the exchange-correlation potential in the Rappe-Rabe-Kaxiras-Joannopoulos 
form.~\cite{Rappe90} The Brillouin zone integration was performed with 
a 15 $\times$ 15 $\times$ 15 {\bf k}-point grid. A kinetic-energy cutoff of 45 Ry 
was employed for the plane-wave expansion of the electronic states.
The experimentally determined lattice parameters and internal atom positions of LaNiPO 
(a = 4.0461 $\AA$, c = 8.100 $\AA$)\cite{Watanabe07} were used. 

To include dynamical correlation effects in the 3$d$ shell of Ni, 
we performed the LDA+DMFT\cite{Held06} calculations for LaNiPO. 
Following the Wannier function projection procedure of Ref.~\onlinecite{Korotin08}, 
we constructed an effective $H_{LDA}$ Hamiltonian and then used it to solve the Dynamical 
Mean-Field Theory (DMFT)\cite{Georges96} self-consistency equations. The $H_{LDA}$ 
Hamiltonian contained 22 bands due to five Ni 3$d$, three O 2$p$, 
and three P 3$p$ orbitals per formula unit, projected in a single energy window that 
explicitly takes into account the hybridization between $p$ and $d$ electrons.~\cite{Korotin08}

The DMFT auxiliary impurity problem was solved by the hybridization function
expansion Continuous-Time Quantum Monte-Carlo method.~\cite{CT} 
The elements of Coulomb interaction matrix were parameterized by $U$ and $J$ 
parameters.~\cite{LAZ95} We used interaction parameters $U=8$ eV and $J=1$ eV for LaNiPO 
similar to the values obtained in~Ref.~\onlinecite{Kunes07}. 
Calculations were performed in the paramagnetic 
state at the inverse temperature $\beta=1/T =$ 20~eV$^{-1}$.
The real-axis self-energy needed to calculate spectral functions was
obtained by the Pad\`e approximant.~\cite{pade}

\section{Results and Discussion} 

\begin {figure}[!t]
\includegraphics[width=0.42\textwidth]{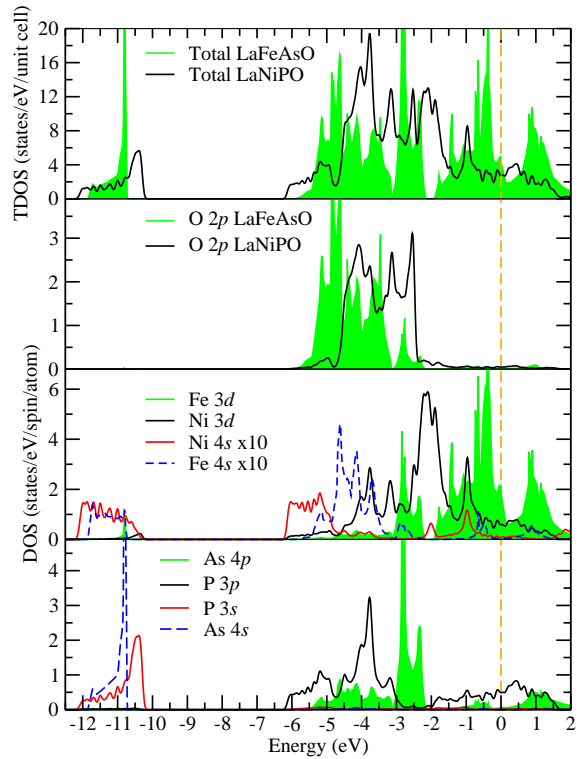}
\caption {(Color online) Total and partial densities of states for LaNiPO (in comparison with LaFeAsO 
from~Ref.~\onlinecite{Anisimov09}) obtained within the density functional theory (DFT) calculations. 
The dashed lines in the Ni, Fe 4$s$, 3$d$ DOS refer to the metal 4$s$~states magnified 
by a factor of 10. La does not have any significant contribution to the valence band, 
and is not shown here.}
\label{fig1}
\end{figure}

The calculated noncorrelated electronic structure of LaNiPO is shown in Fig.~\ref{fig1} in comparison 
with the structure of LaFeAsO.~\cite{Anisimov09} These calculations are in agreement 
with other DOS calculations available to date.~\cite{Lebeque07,Zhang08}
In all cases the far bottom of the valence band (--11 eV) consists of P 3$s$ or As 4$s$. 
The top of the valence band (--2~eV to 0~eV) consists almost solely of metal 3$d$ states in both cases. 
Between --2 and --4 eV there is strong hybridization between the O 2$p$ and Ni 3$d$ states in LaNiPO; 
in LaFeAsO there are far fewer Fe 3$d$ states in this region indicating much weaker hybridization. 
The situation is the same with LaFePO.~\cite{Skornyakov10} LaNiPO also has a reduction 
in metal 3$d$ states and total states at the Fermi level compared to LaFeAsO and LaFePO. 
This may explain why Ni-based superconductors have lower $T_c$ values than FeAs-based 
superconductors. For all compounds the P~3$s$,~3$p$ and As~4$s$,~4$p$ states, respectively, 
occupy the same basic region in the valence band and do not contribute significantly to the Fermi level. 
The La 5$p$ states are identical for both compounds, they have atomic-like character 
and do not contribute to the valence band; they are not shown here. 

\begin {figure}[!t]
\includegraphics[width=0.42\textwidth]{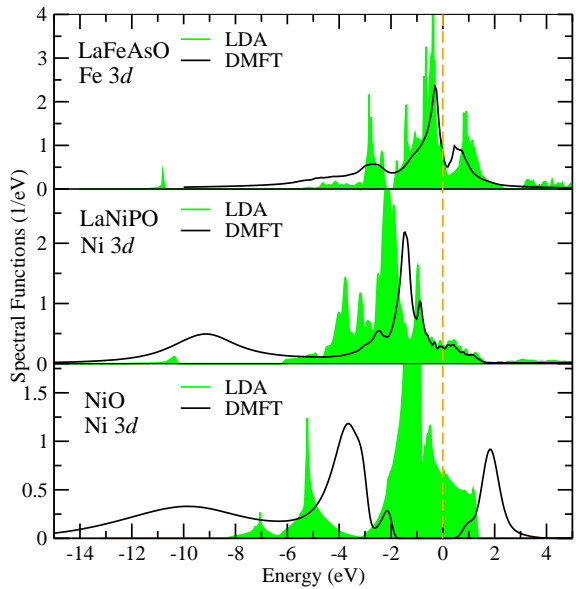}
\caption {(Color online) Densities of states for Ni 3$d$ and Fe 3$d$ orbitals obtained within
DFT (filled areas) and the LDA+DMFT total 3$d$ spectral functions (solid lines). 
Data for LaFeAsO\cite{Anisimov09} and NiO\cite{Kunes07} are given for comparison.}
\label{fig2}
\end{figure}

\begin {figure}[!t]
\includegraphics[width=0.42\textwidth]{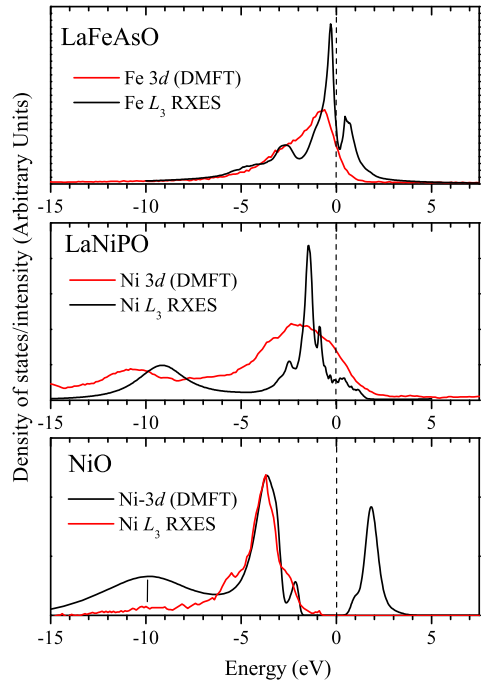}
\caption {(Color online) Comparison of the LDA+DMFT total 3$d$ spectral functions (LaFeAsO from 
Ref.~\onlinecite{Anisimov09} and NiO from Ref.~\onlinecite{Kunes07}) with the resonantly 
excited Ni $L_3$ and Fe $L_3$ X-ray emission spectra.}
\label{fig3}
\end{figure}

The spectral function from the LDA+DMFT calculation for the Ni 3$d$ states of LaNiPO is shown 
in Fig.~\ref{fig2} (middle panel). In the energy interval from --1 to 1~eV near the Fermi 
energy the 3$d$ spectral function is close to the noncorrelated LDA density of states. 
However, below these energies the spectral function is substantially renormalized 
with the formation of a strong peak at --1.5~eV, and the appearance of a lower Hubbard band:
the broad peak centered at --9.1~eV. Thus this picture resembles the LDA+DMFT 
results for LaFeAsO\cite{Anisimov09} (upper panel) only for the energies above --6 eV, 
since in LaFeAsO the lower Hubbard band was not found. In NiO, however, a similar broad peak 
centered at $~$--10 eV is obtained, as shown in the lower panel of Fig.~\ref{fig2}. This lower Hubbard band is evidence for strong correlations,~\cite{Kunes07} and this is a clear indication of strong correlations in LaNiPO,
similar to NiO.~\cite{Kurmaev08a} The comparison of the LDA+DMFT calculation for Ni 3$d$ states 
of LaNiPO and NiO and Fe 3$d$ states of LaFeAsO with the resonantly excited Ni $L_3$ 
and Fe $L_3$ X-ray emission spectra which probe occupied Me 3$d$ states is presented in Fig.~\ref{fig3}. 
The occurrence of the lower Hubbard band in LaNiPO and NiO and its absence in LaFeAsO 
is confirmed by experimental XES spectra.

\begin {figure}[!t]
\includegraphics[width=0.42\textwidth]{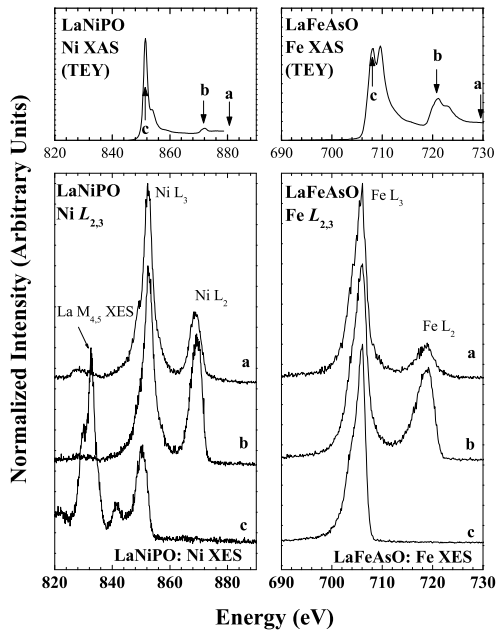}
\caption {Summary of spectra for LaNiPO (left panel) and LaFeAsO (right panel). 
The upper panels show the metal $L_{2,3}$ XAS (in TEY mode), the lower panels 
the resonant and non-resonant metal $L_{2,3}$ XES. The excitation energies 
are indicated by arrows in the XAS plots.}
\label{fig4}
\end{figure}

The soft X-ray metal (Ni, Fe) $L_{2,3}$ spectra are shown in Fig.~\ref{fig4}. The metal $L_{2,3}$ XES 
indicate two main bands separated by the spin-orbit splitting of the metal 2$p$ states. The lower 
intensity high energy band corresponds to the $L_2$ emission line (3$d$,4$s$ $\to$ 2$p$$_{1/2}$ transitions), 
and the higher intensity low energy band corresponds to the $L_3$ emission line (3$d$,4$s$ $\to$ 2$p$$_{3/2}$ 
transitions). The resonant $L_2$ and $L_3$ XES (curves b and c in the bottom panels, respectively) 
have the same basic shape. The lack of resonant features indicates that the spectra primarily 
measuring the partial occupied DOS rather than multiplet or inelastic scattering effects. 
Note that the La $M_{4,5}$ XES appears below the Ni $L_3$ emission line in the resonant Ni $L_3$ spectrum. 
The metal $L_{2,3}$ XAS are presented in the top panels of Fig.~\ref{fig4}. According to dipole selection rules 
($\Delta l$ = $\pm$ 1) they correspond to the excitation of metal 2$p$-core level electrons into unoccupied 
3$d$ states. Unfortunately these spectra can not probe the unoccupied 3$d$ DOS directly because the core-hole 
causes an increased effective nuclear charge distorting the local DOS levels. Further, 
simulating $L_{2,3}$ XAS requires considering multiplet splitting, hybridization, and crystal field effects. 
One such simulation was recently conducted for LaFeAsO in~Ref.~\onlinecite{Kroll08}, to our knowledge 
no similar simulation of LaNiPO exists. Therefore we include the metal $L_{2,3}$ XAS only for completeness. 
Resonantly excited Ni $L_3$ XES of LaNiPO (curve c) shows the presence of La $M_{4,5}$ XES because excitation 
energy in this case is very close to resonant excitation of La $M$-emission spectra.

\begin {figure}[!t]
\includegraphics[width=0.42\textwidth]{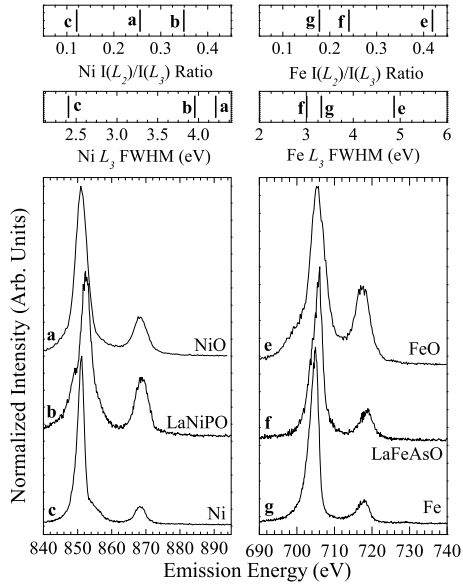}
\caption {Comparison of non-resonant metal $L_{2,3}$ XES of NiO, LaNiPO, and Ni 
(left side panels) and FeO, LaFeAsO, and Fe (right side panels) from Ref.~\onlinecite{Kurmaev08b}. 
The I($L_2$)/I($L_3$) ratios for each system are shown in the top panels, 
and the full width at half maximum (FWHM) of the $L_3$ bands are shown 
in the middle panels. The metal $L_{2,3}$ XES are shown in the bottom panels, 
for easy reference. The I($L_2$)/I($L_3$) ratios were calculated by taking 
the quotient of the integrals of the $L_2$ and $L_3$ bands.}
\label{fig5}
\end{figure}

The ratio of the integral-intensity of the metal $L_2$ and $L_3$ peaks (the I($L_2$)/I($L_3$) ratio) for LaFeAsO 
is roughly the same as that of metallic Fe, and quite different from that of strongly correlated FeO 
(see Fig.~\ref{fig5}, right side, bottom panel). In a free atom, the I($L_2$)/I($L_3$) ratio should 
be equal to 1/2 as the ratio is based solely on the statistical population of the 2$p$$_{1/2}$ 
and 2$p$$_{3/2}$ levels. In metals the radiationless $L_2$$L_3$$M_{4,5}$ Coster-Kronig (C-K) 
transitions greatly reduce the I($L_2$)/I($L_3$) ratio,~\cite{Raghu08} and the I($L_2$)/I($L_2$) 
ratio can be used as a measure for the electron correlation strength of a transition metal compound\cite{Kurmaev05} 
(see Fig.~\ref{fig5}, right side, top panel). The full width at half maximum (FWHM) of the $L_3$ band in LaFeAsO
is again closer to that of metallic Fe than FeO (see Fig.~\ref{fig5}, right side, middle panel). While this does not 
directly prove anything, it suggests that the Fe 3$d$ electronic structure of LaFeAsO may be similar to that 
of metallic Fe. The shape and statistics of the Fe $L_{2,3}$ XES indicate that the Fe 3$d$ states in LaFeAsO 
are not strongly correlated.

In contrast to LaFeAsO, the I($L_2$)/I($L_3$) ratio for LaNiPO is much greater than that of Ni metal, as 
is the FWHM of the LaNiPO $L_3$ band (see Fig.~\ref{fig5}, left side, bottom panel). Indeed, 
the I($L_2$)/I($L_3$) ratio and $L_3$ FWHM for LaNiPO (see Fig.~\ref{fig5}, left side, top and middle panels) 
are rather close to those of correlated NiO. Since the transition metal I($L_2$)/I($L_3$) ratio is over 50\% greater 
in LaNiPO than LaFeAsO, and since in NiO is comparable to FeO in terms of ``correlation strength'',~\cite{Anisimov91} 
the Ni 3$d$ states are more correlated than the LaFeAsO Fe 3$d$ states.

\section{Conclusions} 
We have studied the electronic structure of LaNiPO excited by synchrotron soft X-ray emission 
and absorption spectroscopy and obtained the theoretical spectral functions within 
the combination of local density approximation with Dynamical Mean-Field Theory (LDA+DMFT).  
We conclude that the Ni 3$d$ states of LaNiPO reside deeper in the valence band than 
the Fe 3$d$ states of LaFeAsO. The greater occupation in the metal 3$d$ bands in LaNiPO 
reduces the density of the states at the Fermi level and increases the hybridization with 
O 2$p$ states compared to those in LaFeAsO. Accounting for dynamical correlation 
in the Ni 3$d$ states of LaNiPO results in the renormalization of the states below the Fermi 
energy and the formation of the lower Hubbard band centered at --9 eV, similar to NiO, 
but in contrast to LaFeAsO. The I($L_2$)/I($L_3$) ratio is much higher 
in LaNiPO than in LaFeAsO, indicating the Ni 3$d$ states of LaNiPO have stronger 
electron correlations than the Fe 3$d$ states of LaFeAsO. 

\section{Acknowledgments}
The authors thank J. Kune\v{s} for providing the DMFT code and P. Werner
for the CT-QMC impurity solver used in our calculations. This work was 
supported by the Research Council of the President 
of the Russian Federation (Grant NSH-4711.2010.2), the Russian Science Foundation 
for Basic Research (Projects 08-02-00148, 10-02-00046, and 10-02-00546), 
the Natural Sciences and Engineering Research Council of Canada (NSERC) 
and the Canada Research Chair program, Russian Federal Agency for Science 
and Innovations (Program ``Scientific and Scientific-Pedagogical Training 
of the Innovating Russia'' for 2009-2010 years), grant No. 02.740.11.0217, 
the Dynasty Foundation.

\end {document}